\documentclass[%
 reprint,
 amsmath,amssymb,
 aps,
]{revtex4-2}

\usepackage{graphicx}
\usepackage{dcolumn}
\usepackage{bm}

\newcommand{\abs}[1]{\left| #1 \right|} 
\renewcommand{\d}[2]{\frac{d #1}{d #2}} 
 
\renewcommand\eqref[1]{Eq.\;\ref{#1}} 
\newcommand\numberthis{\addtocounter{equation}{1}\tag{\theequation}} 

\usepackage [english]{babel}
\usepackage [english = american]{csquotes}
\MakeOuterQuote{"}

\usepackage{color}					

\begin{document}

\title{Avalanches impede synchronization of jammed oscillators}

\author{William Gilpin}
 \email{wgilpin@fas.harvard.edu}
\affiliation{%
Quantitative Biology Initiative, Harvard University\\
}%

\date{\today}

\begin{abstract}
Synchrony is inevitable in many oscillating systems---from the canonical alignment of two ticking grandfather clocks, to the mutual entrainment of beating flagella or spiking neurons. Yet both biological and manmade systems provide striking examples of spontaneous desynchronization, such as failure cascades in alternating current power grids or neuronal avalanches in the mammalian brain. Here, we generalize classical models of synchronization among heterogenous oscillators to include short-range phase repulsion among individuals, a property that abets the emergence of a stable desynchronized state. Surprisingly, we find that our model exhibits self-organized avalanches at intermediate values of the repulsion strength, and that these avalanches have similar statistical properties to cascades seen in real-world systems such as neuronal avalanches. We find that these avalanches arise due to a critical mechanism based on competition between mean field recruitment and local displacement, a property that we replicate in a classical cellular automaton model of traffic jams. We exactly solve our system in the many-oscillator limit, and obtain analytical results relating the onset of avalanches or partial synchrony to the relative heterogeneity of the oscillators, and their degree of mutual repulsion. Our results provide a minimal analytically-tractable example of complex dynamics in a driven critical system.

\end{abstract}

\maketitle


\section{Introduction}

Spontaneous synchronization occurs in diverse systems spanning robotic swarms, power grids, neuronal ensembles, and even social networks \cite{gonzalez2014localized,panaggio2015chimera,pikovsky2003synchronization,li2019particle}. Globally-coupled phase oscillators represent a basic, tractable paradigm for understanding synchronization, formulating the phenomenon as the gradual mutual entrainment of individual oscillators due to the compounded effect of continuous weak interactions \cite{pikovsky2003synchronization}. This paradigm stems from pioneering mid-century work by Winfree and Kuramoto, who established the universality of minimal phase oscillator models for understanding synchronization in a wide variety of biological processes \cite{winfree1967biological,kuramoto1984chemical}. Subsequent work has established general results for the onset of synchronization in phase oscillators, and results from these simple, tractable models have been successfully applied to the the study of biological systems spanning from brain waves, to bacterial signalling, to swimming microorganisms \cite{lynn2019physics,cabral2011role,danino2010synchronized,vilfan2006hydrodynamic,gilpin2020multiscale}. However, while the theory of phase oscillators is well-established, in recent years a variety of exotic phenomena have been discovered even in simple model systems of phase oscillators, such as chimera states \cite{kuramoto2002coexistence,abrams2004chimera}, glass-like relaxation mediated by a "volcano" transition \cite{ottino2018volcano}, and oscillation death via broken rotational symmetry \cite{zakharova2014chimera}. These results underscore that even simple abstractions of real-world oscillators can display surprisingly complex dynamics.

Many such discoveries are framed in terms of their effects on the synchronous state, which represents a globally-stable solution of the underlying dynamical equations. However, many real-world oscillator populations, such as neuronal ensembles, exhibit a maximally desynchronous state, in which the phases of individual oscillators become evenly spaced apart on the unit circle \cite{tass2007phase,popovych2005effective}. Such dynamics are achievable, in principle, by introducing negative couplings into standard phase oscillator models \cite{hong2011conformists,tsimring2005repulsive}; alternative approaches include introducing phase offsets or time delays in the interactions among oscillators, or introducing specific pairwise couplings among oscillators that embed them on a complex graph \cite{panaggio2015chimera,nicosia2013remote,omel2008chimera}. The tunability of synchronization in these models introduces the broader question of whether unique phenomena can occur at the critical intermediate point between the synchronous and maximally-desynchronous states.
\\ \indent Here, we investigate this regime by introducing a minimal, tractable generalization of classical oscillator models, which allows a smooth transition between full synchrony and maximal desynchrony. Our model is based on short-range steric interactions observed in swarms and active matter, and it bears resemblance to local inhibitory effects observed in in real-world ensembles of interacting neurons. Our model exhibits a surprising phenomenon consisting of self-organized avalanches, in which the system seemingly approaches synchrony, only to abruptly desynchronize at quasi-random intervals. Despite appearing in a minimal phase model, these avalanches have similar statistical properties and scaling exponents to avalanches observed in real-world oscillator networks, including neuronal ensembles, financial markets, and power grids \cite{beggs2003neuronal,bellenzier2016contagion,motter2013spontaneous}. We trace these avalanches to a critical mechanism driven by competition between large scale attraction of oscillators to the mean field, and small-scale rearrangements, which we recreate using a cellular automaton model inspired by self-organized criticality in traffic. Finally, we solve our model exactly in the many-oscillator limit, and show that avalanches in the system originate in the amplification of "noise" provided by local rearrangements.

%

 
\begin{figure*}
{
\centering
\includegraphics[width=\linewidth]{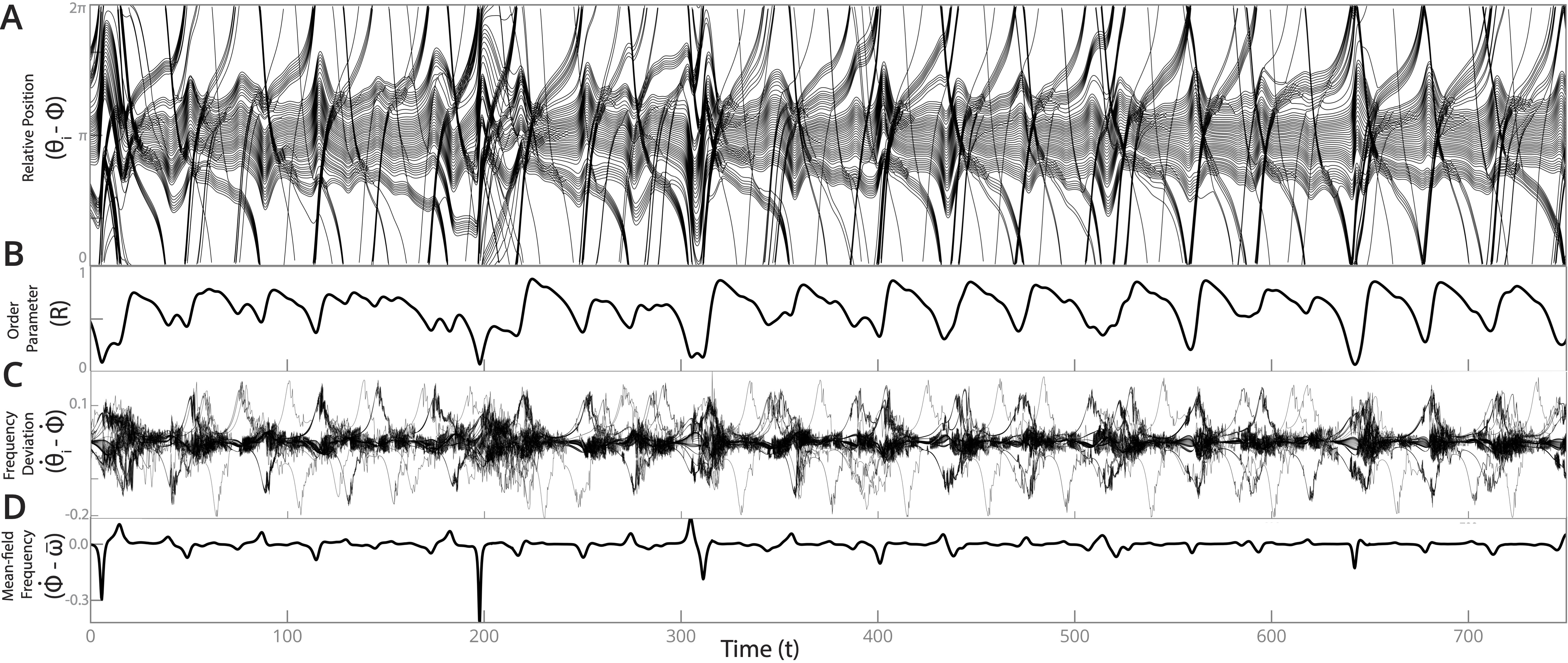}
\caption{
{\bf Transient synchronization and desynchronization in a jammed oscillator ensemble.} (A) A population of $N=60$ oscillators with short-range Gaussian repulsion, visualized in a frame commoving with the mean field. (B) The Kuramoto order parameter $R$; (C) the instantaneous individual frequency relative to the commoving frame; and (D) the mean frequency deviation from the population mean. For this figure and others (except where noted), $K=0.4$, $V_0=0.8$, $\sigma_\omega = 0.2$.
}
\label{avalanche}
}
\end{figure*}

\section{Results}

\subsection{The Kuramoto model with short-range repulsion}

Our model consists of non-identical phase oscillators with global attraction through a classical Kuramoto force; we modify this model by adding an additional term corresponding to short-range repulsion among oscillators with proximal phases,
\begin{equation}
\dot \theta_j = \omega_j + \dfrac1N\sum_{k=1}^N \left( K \sin(\theta_k - \theta_j) - \frac{d}{d\theta_{j}}V(d(\theta_j, \theta_k))\right)
\label{ode1}
\end{equation}
where the interaction $V(.)$ represents a short-range repulsion with amplitude $V_0 \equiv V(0)$ and width $2 \pi/N$; the width scales inversely with $N$ in order to ensure short-ranged interactions. The angular distance function $d(\theta_j, \theta_k)$ corresponds to the signed shortest distance between two angles along the unit circle. The individual oscillator frequencies $\omega_j$ are sampled from a probability distribution $g(\omega_j)$; following previous work we use a Cauchy distribution with mean frequency $\bar\omega$ and width $\sigma_\omega$. Below we obtain comparable results for several choices of $V(.)$.

The repulsive interaction can be seen as an abstraction of steric interactions among individuals in swarms of self-propelled oscillators \cite{li2019particle,levis2017synchronization}. Similar repulsive interactions occur in levitating colloidal particles, which produce collective dynamics reminiscent of those reported here \cite{gogia2017emergent}. Alternatively, in neurons, the repulsion term would correspond to a negative region in each neuron's phase-response curve; such a response could arise due to synaptic depression via local depletion of neurotransmitter reserves \cite{levina2007dynamical,chialvo2010emergent}. 

When $V_0 = 0$, \eqref{ode1} becomes the standard Kuramoto model with global coupling and no phase offsets; this model exhibits solvable dynamics and synchronization when $K$ is large relative to $\sigma_\omega$ \cite{pikovsky2003synchronization}. The onset of synchronization can be readily observed via the dynamics of the Kuramoto order parameter,
\[
R(t) e^{i \Phi(t)} \equiv \sum_{k=1}^N e^{i \theta_k(t)},
\]
which reaches a stable steady-state $R=1$ when $K$ is sufficiently high relative to $\sigma_\omega$ . However, in the case of non-zero repulsion $V_0 > 0$, \eqref{ode1} admits another stable solution when $K=0$ and $\sigma_\omega$ is small, corresponding to equal spacing of oscillators along the unit circle ($R=0$) due to steric interaction. Thus, the relative values of $K$ and $V_0$ parametrize competition between global attraction to the mean field, and local avoidance of clustering.

\subsection{Avalanche dynamics appear at intermediate coupling strengths}

\begin{figure*}
{
\centering
\includegraphics[width=.8\linewidth]{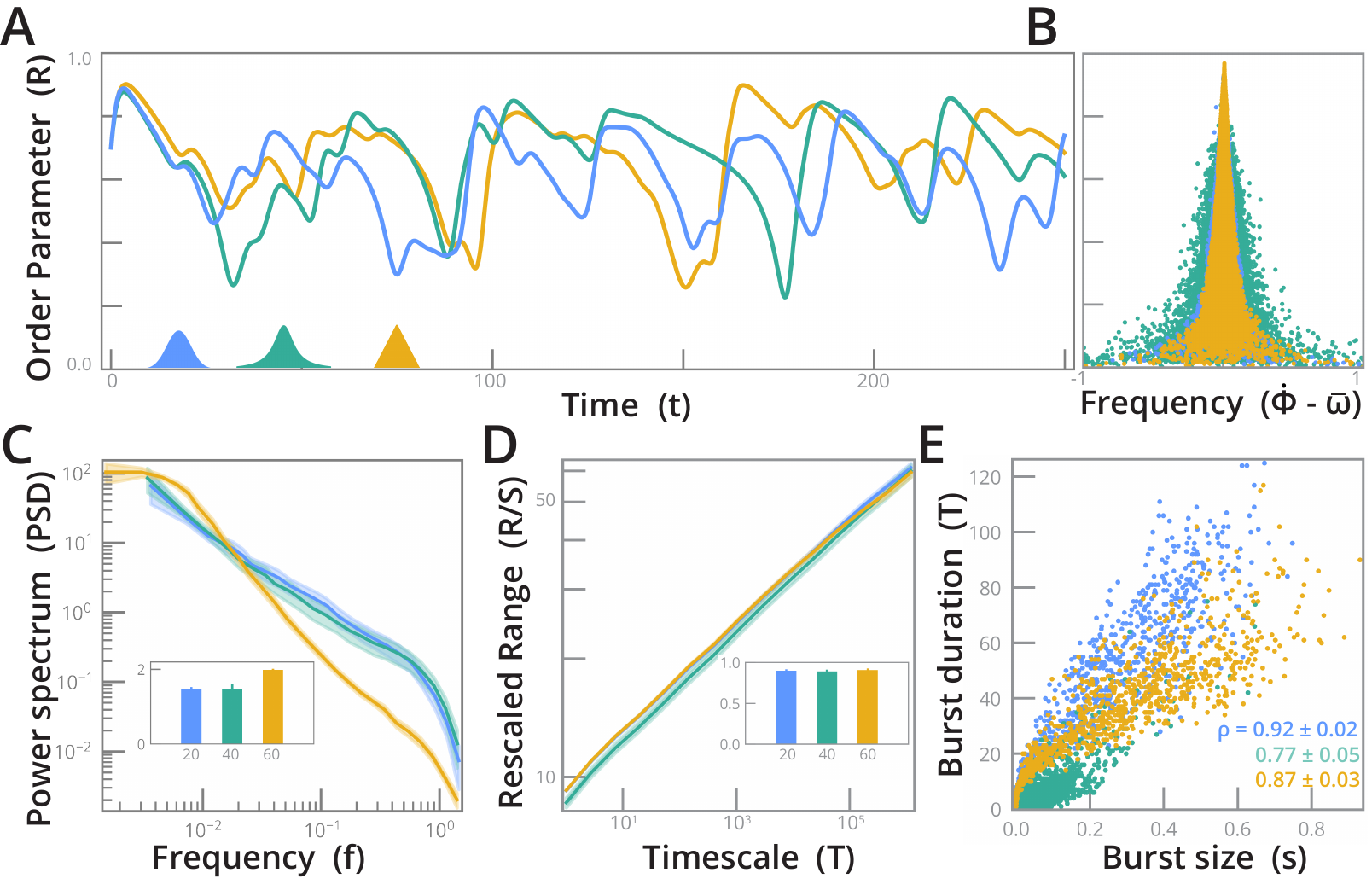}
\caption{{\bf Universal statistical properties of avalanches across different repulsive potentials.} (A) Example dynamics of the order parameter $R$ for short range repulsion given by Gaussian (blue), Cauchy (turquoise), and triangle (yellow) potentials (inset shapes). (B) $R$ versus the deviation between the mean field and the mean intrinsic frequency of the oscillators. (C) The power spectral density, with slopes corresponding to power-law decay exponents (inset). (C) A Hurst plot of the rescaled range versus measurement timescale, a measure of fractality for time series, with slopes corresponding to Hurst exponents (inset). (C) Avalanche duration versus size, annotated by Spearman correlation. All error ranges comprise bootstrapped standard deviations.
}
\label{critical}
}
\end{figure*}

Unexpectedly, at intermediate values of $K$ and $V_0$, the order parameter undergoes irregular oscillations, displaying epochs of rapid synchronization punctuated by gradual desynchronization events (Figure \ref{avalanche}). These fluctuations occur independently of the choice of repulsive kernel $V(.)$; we show results for Gaussian, Cauchy, and triangular potentials in Figure \ref{critical}, although we otherwise focus on Gaussian repulsion for simplicity. Qualitatively, oscillations of the order parameter (and frequency fluctuations of individual oscillators) resemble those of relaxation oscillators; intuitively, the repulsive term in \eqref{ode1} introduces a second timescale into the collective dynamics (the first being the synchronization rate). As in other oscillating systems, this timescale separation introduces the potential for complex dynamics---similar fluctuations arise in relaxation oscillators, due to timescale separation between fast and slow manifolds, as well as in integrate-and-fire neuron models. In our model, the second timescale arises directly from interaction with other oscillators.


These irregular oscillations are particularly striking given the global coupling among units. Self-organized quasiperiodicity has previously been reported in the phases of individual phase oscillators with nonlinear coupling \cite{rosenblum2007self}; in this system, when the coupling $K$ is too low to produce full synchrony, the mean field fails to entrain individual oscillators---however, $R(t)$ remains periodic. The behavior of \eqref{ode1} also differs from partial synchronization observed in chimera states, in which distinct synchronous and desynchronous subpopulations coexist within an ensemble of oscillators \cite{kuramoto2002coexistence,abrams2004chimera,panaggio2015chimera,martens2013chimera}. While chimera states can produce irregular oscillations of $R(t)$, the form and character of chimera states arises specifically from the presence of mixed short- and long-ranged pairwise couplings. In contrast, the number of short-range interactions experienced by an oscillator subject to \eqref{ode1} varies continuously as other oscillators enter and exit its effective repulsive radius.

\subsection{Characterizing the critical mechanism}

We hypothesize that our observed cascades arise from a critical mechanism, due to: (1) the universal long-term statistics of our system, which are invariant to the choice of interaction kernel; (2) the occurrence of cascades at intermediate regimes between fully-dispersed and fully-localized states; and (3) the lack of bistability in \eqref{ode1}. Lending quantitative support to this hypothesis, we observe that the power spectrum of the order parameter $R(t)$ exhibits $1/f^\alpha$ decay, with $\alpha = 1.5 \pm 0.08$ (Fig. \ref{critical}C); additionally, the higher-order order parameters $R_n \equiv \sum_{j=1}^N e^{i n \theta_j(t)}, n > 1$ exhibit more gradual decay at low frequencies, indicating stronger temporal correlations in the clustering of oscillators over long timescales. Similar scaling exponents occur in several systems with self organized criticality, including sandpile models, granular relaxation, and experimental measurements of "stick-slip" friction \cite{bak1987self,jensen198ff,held1990experimental,feder1991self}. Further evidence of criticality comes from the scaled range versus time interval, which shows a straight-line trend indicative of self-similarity over five orders of magnitude (Fig. \ref{critical}D). The slope gives a Hurst exponent $H = 0.91 \pm 0.01\approx 1$ indicating long-time correlations consistent with criticality \cite{kantz2004nonlinear}; experimental work has observed that $H \approx 1$ may indicate criticality even in finite systems \cite{kapiris2004electromagnetic}. Consistent with a critical mechanism, we find strong positive correlation ($0.92 \pm 0.02$, 954 events) between the size of individual drops in $R$ (peak-to-trough amplitude) and the duration of each event (peak-to-peak distance)---a sign of a critical relaxation mechanism driving desynchronization \cite{bak1987self,jensen198ff,friedman2012universal}. 

We do not rule out whether avalanches are transient within our system, although we note that transience would not preclude criticality, because non-conservative systems can be transiently critical. For example, semi-disordered chimera states originally observed in discrete oscillator ensembles were later found to be transient, but very long-lived \cite{wolfrum2011chimera,panaggio2015chimera}. However, we note that when $\sigma_\omega =0$ the avalanches are necessarily transient because \eqref{ode1} can be written as the gradient of a potential \cite{van1993lyapunov}; however, we observe in practice that fluctuations rarely cease when $N>20$ or $\sigma_\omega > 0.01$, introducing the possibility that our phenomenon is either stable or a "supertransient" that scales sharply with system size \cite{wolfrum2011chimera}. In any case, the persistence of avalanches, and the wide range of parameter values over which they occur, suggest that our observed dynamics are an important phenomenon for systems of the form of \eqref{ode1}---especially real-world systems with noise or other driving that could continuously re-trigger cascades.

\begin{figure}
{
\centering
\includegraphics[width=\linewidth]{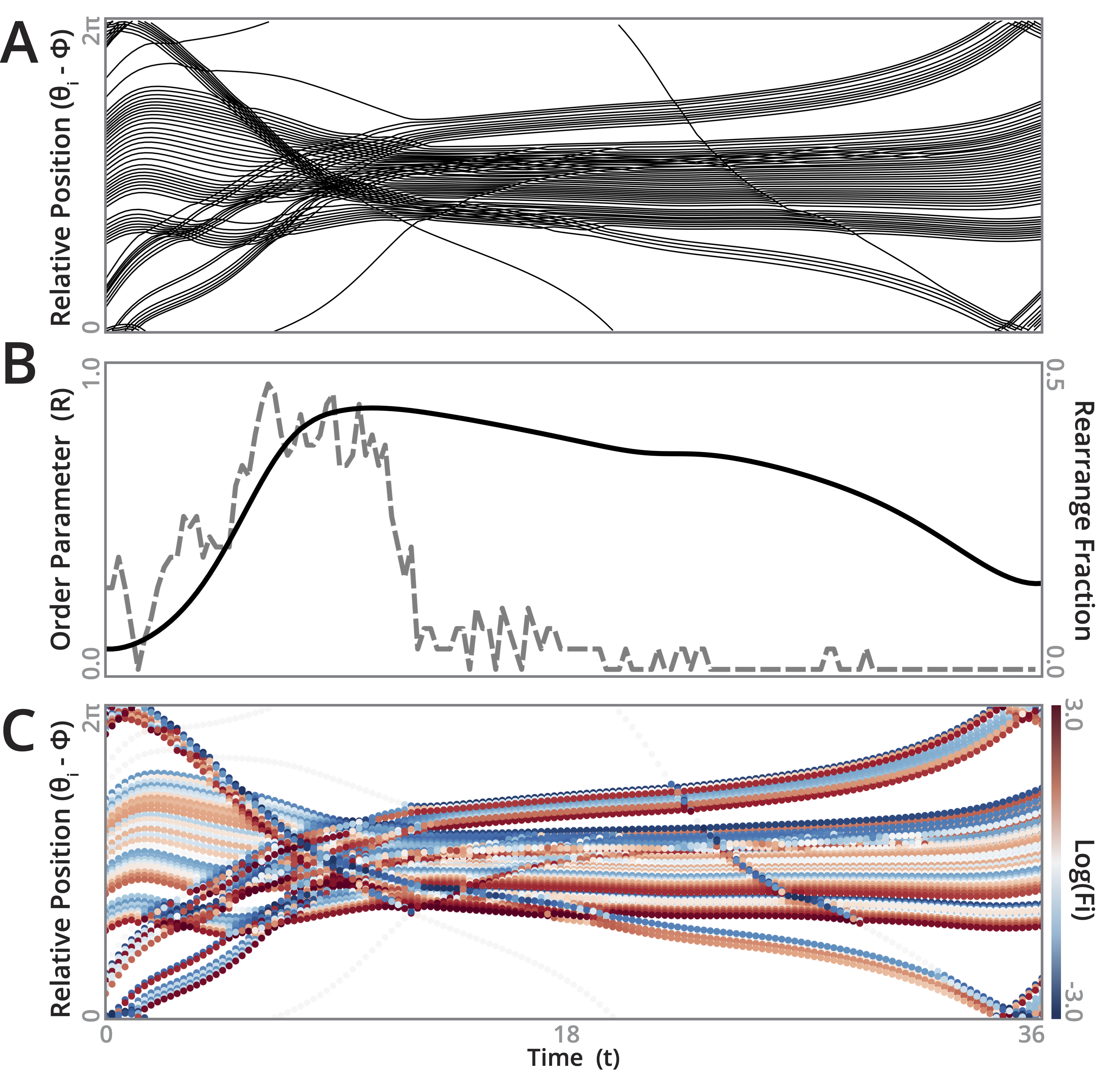}
\caption{{\bf Microscale oscillator rearrangements trigger avalanches.} (A) Positions of oscillators during a typical synchronization/desynchronization cycle. (B) The order parameter $R$ (left axis) and the fraction of rearrangements per timestep (right axis). (C) The oscillator positions colored by net repulsive force $\mathcal F$.
}
\label{mechanism}
}
\end{figure}

Having observed long-term statistical properties consistent with self-organized criticality, we next consider the microscopic mechanism of criticality in our system. Close inspection of a single synchronization-desynchronization cycle reveals that rearrangements drive the onset of avalanches (Figure \ref{mechanism}). Intuitively, attraction of oscillators towards the mean field causes an exponential increase in local density in $\theta$-space. This clustering increases the steric pressure on synchronized oscillators, increasing the probability that two oscillators will overlap sufficiently to either exchange positions, or exert a joint force on a third oscillator---thereby triggering a cascade of oscillator rearrangements and transient desynchronization. We quantify this effect by computing the net repulsive force acting an oscillator $j$, due the cumulative effect of other oscillators falling within its radius of interaction,
\[
\mathcal F_j = -\sum_{k=1}^N  \frac{d}{d\theta_{j}} V(\theta_j, \theta_k)
\]
which is proportional to the gradient of the local density of oscillators: an oscillator with an equal number of neighbors falling on its right and left sides experiences zero net force. Figure \ref{mechanism}C overlays $\mathcal F_j$ on a typical cycle, and Figure \ref{mechanism}B shows the "rearrangement fraction," the fraction of oscillators that exchange positions at each timepoint. During a single timestep, if all oscillators exchange and permute positions along the unit circle due to rearrangements, this quantity will be $1$; if no oscillators exchange positions, then the rearrangement fraction will be $0$. We observe that, during synchronization, brief peaks in the net force appear and then relax, due to oscillators having sufficient space to rearrange and maximize their spacing. However, as the degree of synchronization increases, oscillators become less likely to have space to freely rearrange, causing repulsive forces to aggregate within the cluster. Eventually, large-scale rearrangement occurs, triggering a cascade of rearrangements that break apart the cluster---visible as a sustained period of non-zero net repulsive force, position exchanges, and dispersion along the unit circle.

Given our observation that criticality arises due to competing aggregative and dispersive processes occurring over widely-separated timescales, we seek to recapitulate our findings in a minimal model that rules out potential alternative mechanisms \cite{boffetta1999power}. The attraction of oscillators towards the mean field qualitatively resembles directed flow in classical traffic models, with steric interactions among oscillators seeding "jams" due to excluded volume effects. We start with the classic Nagel-Schreckenberg cellular automaton model for traffic on a periodic domain \cite{nagel1992cellular}. In this model, $N$ cars are distributed among $M$ sites, represented by a length $M$ string with site values $0$ (unoccupied) or $1$ (occupied, without overlaps). In each timestep, cars at each occupied site move forward a number of units determined by their current velocity, and then update their velocity using various acceleration rules (such as random acceleration or braking). A central feature of the model is that a car's velocity is bounded by the number of empty sites preceding it, thus preventing passing---a constraint that triggers the formation of large-scale jams at high densities $N/M \rightarrow 1$, and which allows the system to exhibit self-organized criticality when the density is sufficient for jams to transiently form and then break apart \cite{nagel1995emergent}. Here, we modify this model in two ways: (1) we impose an acceleration rule that depends, in part, on the distance of each car from all other cars, and (2) if the net repulsive force on a given car reaches a critical threshold, then the car accelerates at a rate depending on the magnitude and direction of its net force. The latter rule comprises a one-dimensional toppling mechanism for large jams, as occur in classical sandpile models of self-organized criticality \cite{bak1987self}. We describe the model in greater detail in the appendix. 

\begin{figure*}
{
\centering
\includegraphics[width=0.7\linewidth]{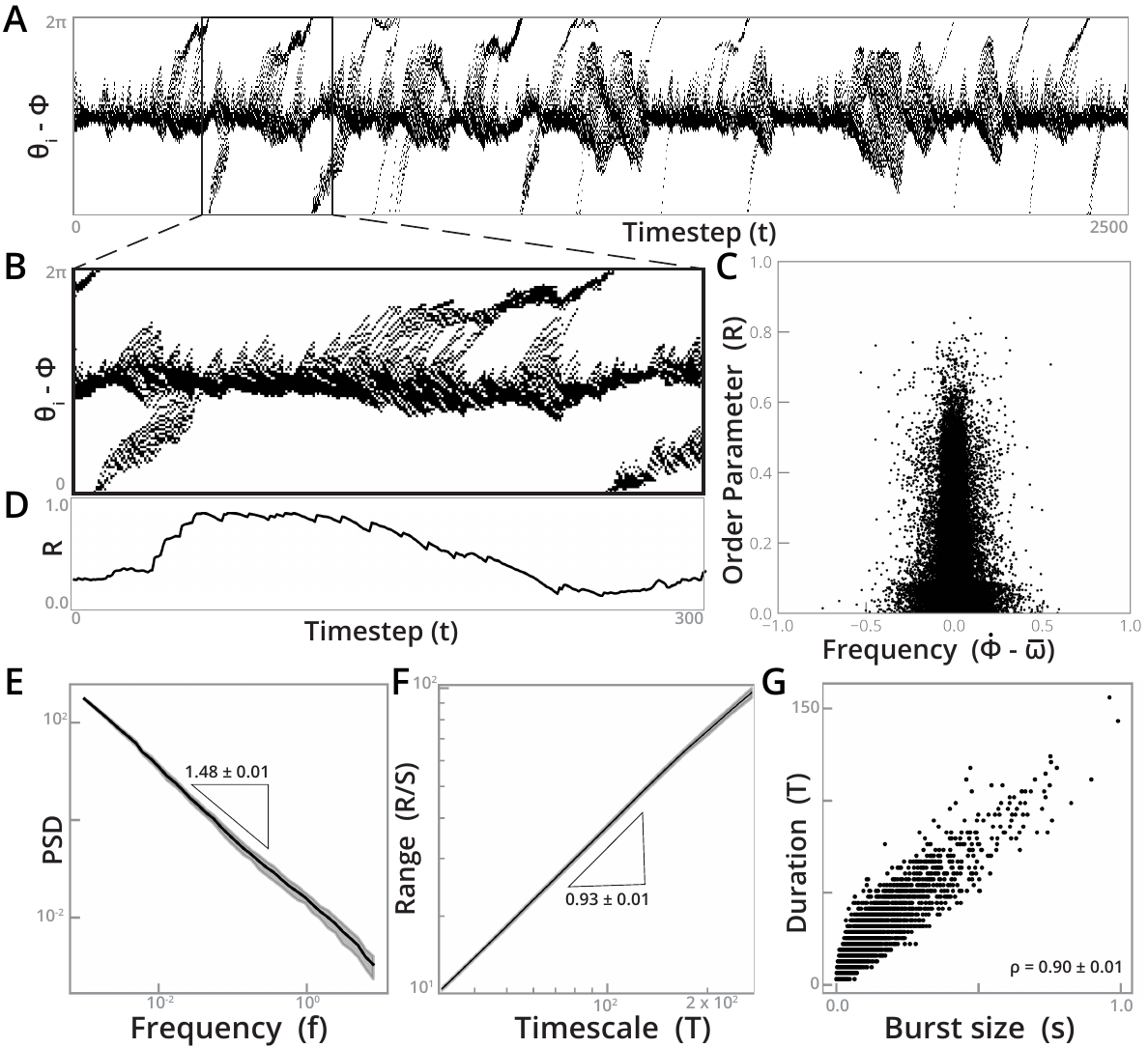}
\caption{{\bf A cellular automaton traffic model recreates key features of the dynamics.} (A) The positions of $60$ cars in the commoving frame of a toppling cellular automaton traffic model. (B) A magnified single avalanche event, and (C) the order parameter $R$. (D) $R$ versus the deviation between the mean field and the mean intrinsic frequency. The (E) power spectral density and (F) Hurst scaling plot, with power law exponents annotated, and (G) the topple duration versus size, annotated with the Spearman correlation coefficient. Error ranges comprise bootstrapped standard deviations.
}
\label{traffic}
}
\end{figure*}

We observe that our modified traffic automaton replicates the irregular build-up and breakup of synchronized states that we observe in the continuous-time oscillator dynamics (Figure \ref{traffic}). Moreover, the statistical properties of the two models are comparable: the power spectral density of $R$ (calculated by assuming the cars travel on the unit circle) displays $\alpha = 1.48 \pm 0.01$, while the raw time series exhibits $H = 0.93 \pm 0.01$. The similarity between the oscillator and traffic models underscores the central role of force build-up and rearrangement in determining our observed avalanche dynamics, despite qualitative differences between the two models. 

\subsection{A continuum model maps the critical regime}

\begin{figure*}
{
\centering
\includegraphics[width=0.6\linewidth]{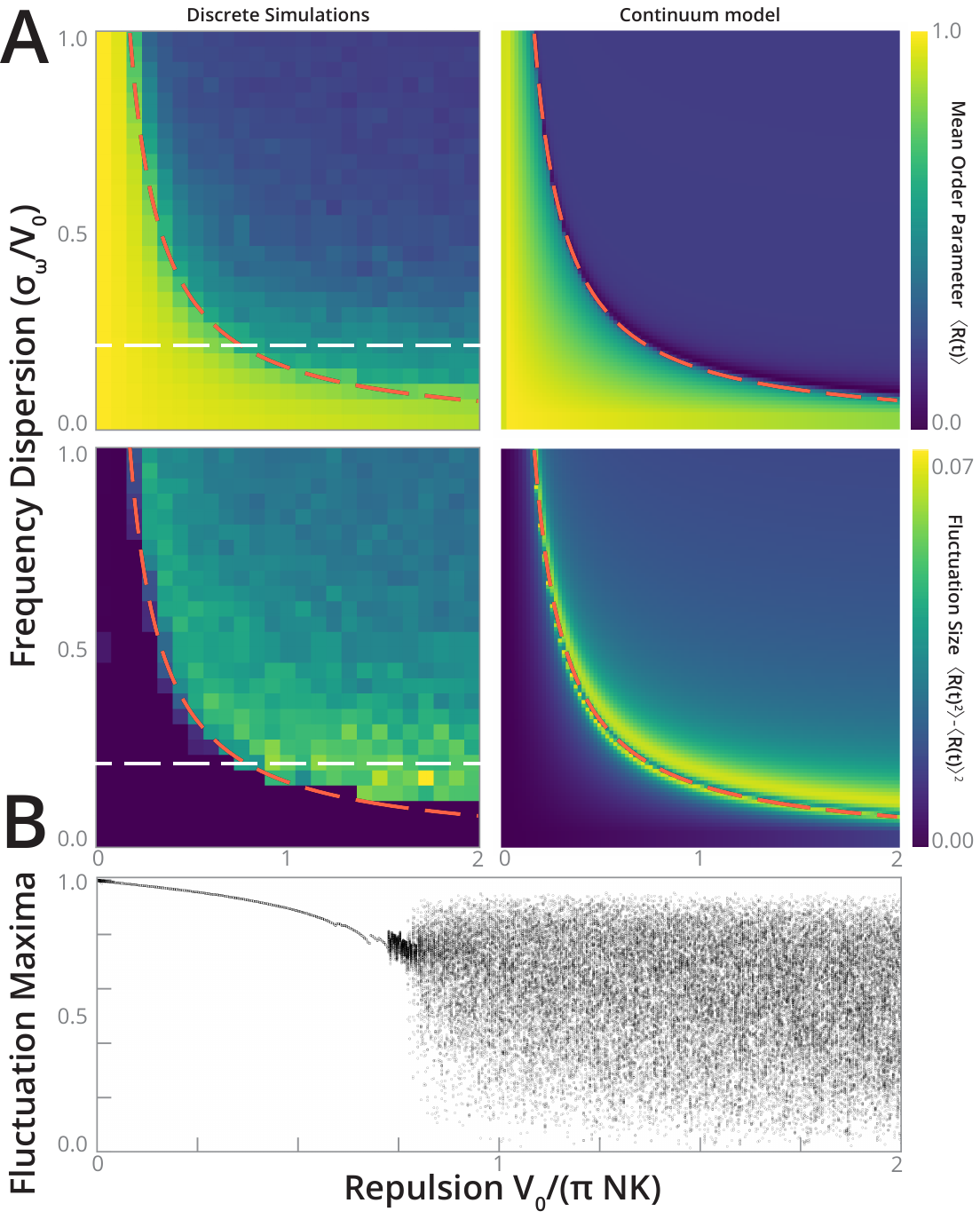}
\caption{
{\bf A stochastic continuum model summarizes the phase space of avalanches.}  (A) The empirical phase space of fluctuations in $R$ as the standard deviation of oscillator frequencies and relative repulsion strength are varied, and the corresponding phase space for the continuum model. The analytic stability boundary for the continuum model is overlaid (red dashes).  White horizontal line indicates a cross section of the phase space plotted in (B), which shows a bifurcation diagram comprising the extrema taken by $R$ during steady-state fluctuations as $V_0/(\pi N K)$ increases.
}
\label{model}
}
\end{figure*}

That rearrangements trigger cascades indicates the role of inter-oscillator repulsion in amplifying local events into global cascades. Avalanches therefore are the result of \eqref{ode1} becoming extremely sensitive to noise (here provided by rearrangements) at intermediate values of $V_0/K$. In order to better understand this critical mechanism, we next seek to solve the repulsive oscillator model analytically in the continuum (large $N$) limit. In this limit, the system comprises an "oscillator fluid" with the dynamical equation
\begin{equation}
\partial_t f + \partial_\theta (v f) = 0,
\label{hydro}
\end{equation}
where $f(\theta, \omega, t)$ is the probability density of oscillators at position $\theta$ along the unit circle. The force is given by a continuum analogue of \eqref{ode1},
\begin{equation}
v = \omega + K \int \int f\,\sin(\theta' - \theta) d\theta' d\omega- (1/N)(V_0 + \xi(t))\, \partial_\theta f.
\label{velocity}
\end{equation}
The first two term in this expression follow directly from the Kuramoto model \cite{ott2008low}. The remaining term proportional to $\partial_\theta f$ follows from the continuum limit of the repulsive interaction (see appendix for derivation). Similar density gradient terms appear in continuum versions of critical sandpile models \cite{gil1996landau}; here, the term helps stabilize the uniform distribution $f = 1/(2\pi), R = 0$. The noise term $\xi(t)$ triggers cascades; rearrangements are seeded by uncorrelated stochastic events $\langle \xi(t) \xi(t') \rangle = \xi_0 \delta(t - t')$, with amplitude proportional to the local oscillator density. We discuss noise in greater detail below.

In order to solve for the dynamics of $R(t)$ subject to \eqref{hydro}, we impose a form for $f(\theta)$ using the Ott-Antonsen ansatz \cite{ott2008low,bick2020understanding}, which assumes a Fourier representation of the density distribution,
\[
f(\theta) = \dfrac{g(\omega)}{2 \pi} \left(1 + \sum_{n=1}^\infty a^n e^{i n \theta} + \text{c.c.}\right).
\]
We insert this equation in \eqref{hydro}, and we assume that $g(\omega)$ is given by a Cauchy distribution of width $\sigma_\omega$. The resulting dynamical equation reveals that the full dynamics are captured by those of the order parameter,
\begin{align*}
&\dot R =  -\sigma_\omega R(t) -\dfrac{R(t)}{2 \pi}   \bigg(    \pi K (R(t)^2 - 1) + \dfrac{P(R(t))}{N}(V_0 + \xi(t)) 		\bigg), \\
&P(R(t))  \equiv 1 + R(t)^2 + R(t)^4 + R(t)^6 + R(t)^8
\numberthis
\label{oder}
\end{align*}
where $R(t) e^{i \Phi(t)} \equiv \int e^{i\theta'}  f(\theta') d\theta'$. In the zero-noise limit ($\xi = 0$), \eqref{oder} admits two physically-meaningful solutions,
\begin{equation}
R=0, \quad
R \approx \dfrac{\sqrt{2}}{4}  \sqrt{\sqrt{u^2 +16 \left(u-1\right) -16 \left(\dfrac{2 \pi  \sigma_\omega}{V_0}\right)} - u   },
\label{rsol}
\end{equation}
where $u \equiv \pi K/V_0$. The approximate nonzero solution closely matches the exact solution expressed in terms of the root of the high-degree polynomial $P(R)$ (see appendix). Analysis of the eigenvalues associated with these solutions reveals that they exchange stability via a transcritical bifurcation at  $V_0 = \pi N (K - 2\sigma_\omega)$. Discrete oscillator simulations across various $\sigma_\omega$ and $V_0/K$ show general agreement with \eqref{rsol} below this critical value (Figure \ref{model}); in this regime the dynamics converge to a partially-synchronized state. However, as $V_0 \rightarrow  \pi N (K - 2\sigma_\omega)$, the dynamics of the discrete simulations undergo an abrupt transition to either stable maximal desynchronization $R=0$ when $\xi = 0$ or avalanche dynamics when $\xi > 0$.

Together, \eqref{hydro} and \eqref{velocity} constitute a damped Burgers' equation with nonlocal forcing; previous studies have demonstrated that Burgers' and Kardar-Parisi-Zhang models can exhibit critical dynamics when driven by noise \cite{diamond1995dynamics,szabo2002self}. We therefore seek to recreate avalanches observed in our discrete simulations by introducing noise ($\xi(t) > 0$) into \eqref{velocity}. Analysis of the Ott-Antonsen ansatz and the resulting dynamical equations shows that the simplest way that noise can affect the dynamics of $R(t)$ is through a term of the form $\xi(t)\partial_\theta f$, as appears in \eqref{velocity}; other approaches, such as additive or multiplicative noise in \eqref{velocity}, do not affect the dynamics of $R(t)$ (see appendix). We derive a stochastic differential equation $dR = F(R) dt + \xi_0\, G(R) dW$, where the deterministic term $F(R)$ is given by the right hand side of \eqref{oder}, and $G(R) = - (R(t)/2 \pi) P(R(t))$. We simulate this Stratonovich process for a long duration with different pairs $V_0/K$ and $\sigma_\omega$, discard a transient period, and then report the long-timescale statistics of $R(t)$ in Figure \ref{model}. We record the mean $R$ to facilitate comparison to \eqref{rsol}, and we record the variance in $R(t)$ to measure the presence and degree of fluctuations in $R$ due to avalanching. We observe that key features of the phase space exactly match between the discrete simulations and theoretical results, despite the relative simplicity of our rearrangement model. Moreover, the analytic stability boundary $V_0 = \pi N (K - 2\sigma_\omega)$ determines the onset of fluctuations (Figure \ref{model}, dashed red trace), which only occur when the maximally-desynchronous state $R = 0$ is stable Thus, while our noise term does not explicitly model small-scale details of rearrangements, it shows that they essentially act as a source of random variation that is amplified by the dynamics at intermediate values of $V_0/K$. Differences between the theoretical and observed results are most apparent in the shape and decay of the threshold between partial synchronization and avalanches (or maximal desynchony when $\xi = 0$); a more detailed microscopic rearrangement model would likely resolve these discrepancies. We note that beyond capturing the general form of phase space, the continuum model captures non-trivial properties of our observed avalanches, such as an uptick in the relative amplitude of $R$ fluctuations when $\sigma_\omega \approx 0.1 - 0.3$.

\section{Discussion}

We have shown that a minimal generalization of classical synchronization models produces unexpectedly rich dynamics, mirroring those of systems exhibiting self-organized criticality and avalanches. We anticipate potential applications of our model to understanding avalanche-like dynamics seen in real-world oscillator networks, ranging from power grids \cite{motter2013spontaneous}, to financial markets \cite{bellenzier2016contagion}, to neuronal circuits in the brain \cite{beggs2003neuronal}. Our observed dynamics exhibit power law scaling of power spectral density, multifractality, burst build-up, and other characteristic features of self-organized critical phenomena \cite{mallinson2019avalanches}; for example, neuronal avalanches have previously been reported to exhibit power spectral densities with critical exponents between $1$ and $2$, and Hurst exponent $\sim\!0.7$ \cite{friedman2012universal,palva2013neuronal,fontenele2019criticality,pu2013developing,jannesari2020stability}. However, while the Kuramoto model has frequently been used as a minimal model of neuronal synchronization \cite{lynn2019physics,cabral2011role}, further study is needed in order to determine whether neurons undergoing avalanches can be mapped onto our modified Kuramoto model---particularly because real-world neuronal networks have sophisticated spatial structure and highly nonlinear interactions \cite{lynn2019physics,chialvo2010emergent}. Nonetheless, our results suggest that even neurons with a simple interaction scheme (all-to-all coupling) could produce avalanches when individual neurons have a negative region in their phase-response curve \cite{dodla2017effect}, a property that was recently shown to induce complex dynamics in experimental oscillator networks \cite{cualuguaru2020first}. We speculate that, in our model, short-range phase repulsion acts analogously to synaptic depression, in which neurons inhibit one another over short time periods due to local depletion of a neurotransmitter---a mechanism that has previously been shown to be sufficient to produce avalanches with comparable statistical properties to our system \cite{levina2007dynamical,chialvo2010emergent}. More broadly, our work provides a minimal example of globally-coupled oscillators that produce dynamics poised between synchrony and disorder, illustrating how critical dynamics can increase the sensitivity of real-world oscillator ensembles to noise and external perturbations.

\section{Acknowledgments}

We thank Daniel Forger and Iryna Omelchenko for their comments on the manuscript. W. G. was supported by the NSF-Simons Center for Mathematical and Statistical Analysis of Biology at Harvard University, NSF Grant DMS 1764269, and the Harvard FAS Quantitative Biology Initiative.

\bibliography{avalanche_cites}

%
\renewcommand{\thetable}{S\arabic{table}}
\setcounter{table}{0}
\renewcommand{\thefigure}{S\arabic{figure}} 
\setcounter{figure}{0}
\renewcommand{\theequation}{A\arabic{equation}}
\setcounter{equation}{0}
\renewcommand{\thesubsection}{\Alph{subsection}}
\setcounter{subsection}{0}
\newpage





%

\begin{figure}
{
\centering
\includegraphics[width=\linewidth]{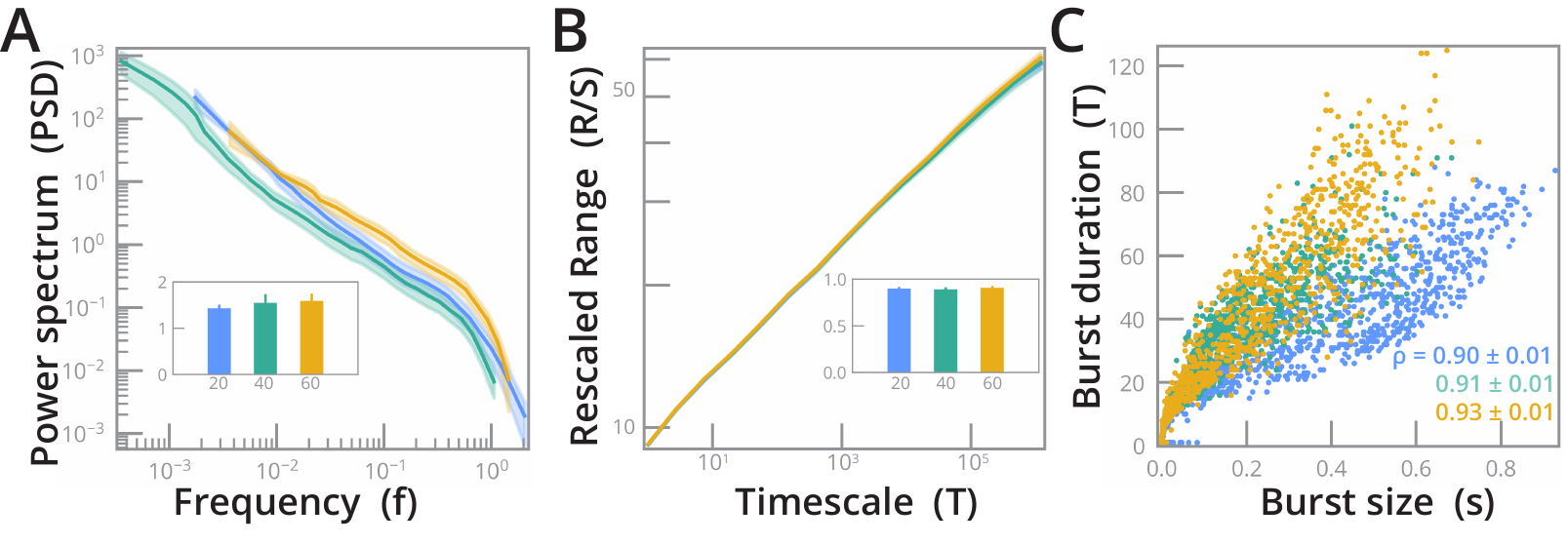}
\caption{
{\bf Dependence of avalanche dynamics on the number of oscillators.}  (A) The power spectral density, with slopes corresponding to power-law decay exponents inset, for $R$ from simulations of $20$ (blue), $40$ (turquoise), and $60$ (yellow) oscillators with Gaussian repulsion. (B) A Hurst plot of the rescaled range versus measurement timescale, a measure of fractality for time series, with slopes corresponding to Hurst exponents inset. (C) Avalanche duration versus size, with Spearman correlation inset. All error ranges correspond to bootstrapped standard deviations.
}
\label{scalingN}
}
\end{figure}

\section{Materials and Methods}

\subsection{Numerical simulations and analysis techniques}

Python 3 code used to simulate the repulsive oscillator model is available on GitHub: \url{https://github.com/williamgilpin/kurep_paper}

All potential functions are defined on the range $[0, 2 \pi]$, with width parameters that scale inversely with $N$. For the triangular potential function, the derivative (comprising two square functions) is calculated and smoothly approximated using a superposition of logistic functions. The qualitative appearance and properties of the avalanches do not vary as the smoothness of the logistic approximation is varied.

Numerical integration is performed using the method of lines. Several solvers and timesteps were tested and compared (see Section \ref{numerics} below); we found best results using a standard LSODA solver as interfaced by {\verb scipy.integrate.solve_ivp }. Before integrating, we compile the right hand side of the dynamical equation using \texttt{numba} (a Just-in-Time compiler for Python). 

All power law fits, error ranges, and visualizations are calculated using recommended best-practice techniques based on the cumulative distribution function \cite{clauset2009power}, as implemented in the \texttt{powerlaw} Python package \cite{alstott2014powerlaw}. We use the default methods recommended for fitting power laws: optimal $x_{min}$ and $x_{max}$ are found by computing the minimum Kolmogorov-Smirnov distance between the data and the fit, and we confirm the quality of our model by performing a goodness-of-fit test between our data and the power law model, which we compare against stretched exponential and lognormal distributions \cite{clauset2009power}. All fits were repeated across $500$ time series in order to generate error bars.

Hurst exponents and scaling plots are produced using the standard $R/S$ algorithm\cite{mandelbrot1969robustness} with RANSAC fitting, as implemented in \texttt{nolds} \cite{scholzel2019}. Observed $R/S$ scalings and calculated Hurst exponents were confirmed by comparing against surrogate time series with shuffled data. Equivalent scaling exponents were observed using detrended fluctuation analysis; however, we favor reporting Hurst exponents due to the stationarity of our time series \cite{peng1994mosaic}, and the relative interpretability of Hurst exponents for multifractal time series.

When calculating avalanche size and duration, we use a numerical peak-finding algorithm to find all extrema in a time series of $R(t)$. We define the size of an event as the difference between a maximum and the next minimum. We define the duration of an avalanche as the number of timepoints that elapse between two successive minima. We compute Spearman correlations between these two quantities, and we generate error ranges for the reported correlation coefficients by bootstrapping $500$ randomly-chosen subsets of the time series.

\section{Continuum limit of the summed repulsive interactions}

The dynamical equation for $\dot\theta_j$ contains the term
\begin{equation}
\dfrac1N\sum_{k=1}^N \frac{d}{d\theta_{j}}V(\theta_j, \theta_k)
\label{repel_term}
\end{equation}
Where $V(\theta_j, \theta_k) = V(D(\theta_j, \theta_k))$, and $D$ is a distance function on the unit circle that gives the shortest signed angular distance along the unit circle (wrapping around at $\pi$), $D(\theta_j, \theta_k) = \mod\!\!(\theta_j - \theta_k - \pi,\, 2\pi) - \pi$.

We take the continuum limit of \eqref{repel_term}, and we assume that $V$ decays sufficiently quickly that $V(\theta_j, \theta_k) \rightarrow 0$ as $\abs{\theta_j - \theta_k} \rightarrow \pi$, and so $V(\theta_j, \theta_k) = V(\theta_j -  \theta_k)$. \eqref{repel_term} then has the form of a convolution,
\begin{equation}
\int f(\theta') \frac{d}{d\theta}V(\theta - \theta') \,d\theta'
\label{continuum_limit}
\end{equation}
where $f(\theta')$ is a probability distribution of phases on the interval $[0, 2\pi]$. Next, we note the following property of convolutions,
\[
\d{}{t} (f \ast g) = f \ast \d{g}{t} = \d{f}{t} \ast g
\]
we therefore exchange the derivatives in \eqref{continuum_limit},
\begin{equation}
\int \frac{d f}{d\theta} V(\theta - \theta') \,d\theta'
\label{swap}
\end{equation}
We assume that, in the continuum limit $N \rightarrow \infty$, $V(\theta - \theta') \rightarrow \delta(\theta - \theta')$, which is true as long as $V$ is compact and has a width parameter that decreases as $N$ increases (as we require for $V(.)$ in the main text). For example, if $V$ corresponds to a Gaussian distribution $V = G_{\mu=0,\sigma}$, then this is equivalent to the condition $\sigma = \sigma_0 / N$ with $\sigma_0$ constant. In this case, \eqref{swap} simplifies to $\partial_\theta f$. The full continuous-time dynamical equation thus becomes
\begin{equation}
\partial_t f + \partial_\theta (v f) = 0,\quad
v = \omega + K \int f(\theta')\sin(\theta' - \theta) d\theta'- (V_0/N) \partial_\theta f.
\label{fulleq}
\end{equation}
The factor of $N$ on the repulsive term appears because we defined the discrete system in the main text extensively: as the number of oscillators $N$ increases, the maximum force attainable by overlapping oscillators also increases by a factor of $N$.

\section{Ott-Antonsen reduction without noise}
\label{appendix_oa_no_noise}

The dynamics of a continuous distribution of oscillators in the absence of noise is given by
\begin{equation}
\partial_t f + \partial_\theta (v f) = 0,
\label{supp_hydro}
\end{equation}
where
\begin{align*}
v(\theta, \omega, t) = \omega &+ K \int_{-\infty}^\infty \int_0^{2 \pi} f(\theta', \omega, t)\sin(\theta' - \theta) d\theta' d\omega \\
&- \dfrac{V_0}{N} \partial_\theta f(\theta, \omega, t).
\numberthis \label{supp_vel}
\end{align*}
We define the order parameter, $z(t)$,
\[
z(t) = \int_{-\infty}^\infty \int_0^{2 \pi}  f(\theta', \omega, t) e^{i \theta'} d\theta' d\omega,
\]
for which \eqref{supp_vel} reduces to the form
\[
v(\theta, \omega, t) = \omega + \dfrac{K}{2 i}\left(		z e^{- i \theta} - \bar{z}e^{i \theta}	\right) -  \dfrac{V_0}{N} \partial_\theta f(\theta, \omega, t).
\]

Following previous applications of the Ott-Antonsen reduction \cite{ott2008low,abrams2008solvable,panaggio2015chimera,martens2009exact,kotwal2017connecting}, we express the angular distribution of oscillators in terms of its Fourier series expansion
\begin{equation}
f(\theta, \omega, t) = \dfrac{g(\omega)}{2 \pi} \bigg(	1 + \sum_{n=1}^\infty a_n(\omega, t) e^{i n \theta}	+ \bar{a}_n(\omega, t) e^{-i n \theta}			\bigg).
\label{supp_oaf}
\end{equation}
where $g(\omega)$ is the distribution of frequencies across the population, $g(\omega) = \int_0^{2\pi} f(\theta', \omega, t) d\theta'$.

The Ott-Antonsen ansatz is a low-order closure for this moment series, $a_n(\omega, t) = a(\omega, t)^n$, $\bar{a}_n(\omega, t) = \bar{a}(\omega, t)^n$. Inserting this ansatz into \eqref{supp_oaf}, and then inserting the result into \eqref{supp_hydro}, results in the equation
\begin{align*}
\dot a(\omega, t) = -\dfrac{a(\omega, t)}{2 \pi}\bigg(2 \pi i \omega - \pi K   &+ \pi K \abs{a(\omega, t)}^2 	\\
&+ (V_0/N) P(\abs{a(t)}	\bigg). \numberthis \label{oa_a}
\end{align*}
with $P(q) \equiv  1 + q^2 + q^4 + q^6 + q^8$. Using the definition of the order parameter, we perform the substitution
\[
z(t) = \int_{-\infty}^\infty g(\omega) \bar{a}(\omega, t) d\omega,\quad \bar{z}(t) = \int_{-\infty}^\infty g(\omega) a(\omega, t) d\omega.
\]
We next choose $g(\omega)$ to be a Cauchy distribution with zero mean $g(\omega) = (\sigma_\omega/\pi)(1/(\omega^2 + \sigma_\omega^2))$. This yields the relation $z(t) = \bar{a}(-i \sigma_\omega, t), \bar{z}(t) = a(-i \sigma_\omega, t)$.

We next apply the substitutions $a(-i \sigma_\omega, t) = R(t) \exp(-i \Phi(t))$,  $\bar{a}(-i \sigma_\omega, t) = R(t) \exp(i \Phi(t))$, $\abs{a(t)} = R(t)$, The real and imaginary parts of the resulting dynamical equation yield the separate dynamics of $\dot R(t), \dot \Phi(t)$
\begin{align*}
\dot R(t) &=  -\sigma_\omega R(t) -\dfrac{R(t)}{2 \pi}   \bigg(    \pi K (R(t)^2 - 1) + (V_0/N) P(R(t))		\bigg) 	\\
\dot \Phi(t) &= \omega
\numberthis
\label{supp_oadyn}
\end{align*}
with high degree polynomial $P(R(t)) \equiv  1 + R(t)^2 + R(t)^4 + R(t)^6 + R(t)^8$. Thus, regardless of the value of the order parameter $R(t)$, the oscillators find a state where all co-rotate at their common frequency $\omega$.

The steady-state values of the order parameter may be calculated by substituting $\dot R(t) = 0$ into \eqref{supp_oadyn}
\begin{equation}
0 = \pi (2 \sigma_\omega -K) + K \pi R(t)^2+ (V_0/N) P(R(t))	
\label{supp_oadyn_ss}
\end{equation}

The steady-states of this equation, as well as the eigenvalues defining their stability, may be solved exactly using existing formulae for the roots of polynomial equations; however, the form of these solutions is not concise, and so we do not reprint them here. These exact solutions predict $0 < R(t) < 1$ as $t \rightarrow \infty$ when $V_0 < \pi N (K - 2\sigma_\omega)$, and that $R(t) = 0$ as $t \rightarrow \infty$ when $V_0 >\pi N (K - 2\sigma_\omega)$. Thus, as the relative strength of the repulsive interaction increases, the steady-state value of the order parameter (and thus degree of synchronization) decreases until reaching zero at $V_0 = \pi N (K - 2\sigma_\omega)$. As the repulsive interaction further increases, the steady state remains at zero. Thus, at $V_0 = \pi K N$ the system undergoes a transcritical bifurcation.

To better illustrate the behavior of the full solution, we note that a concise approximate solution may be found by noting that $R(t)^2 + R(t)^4 + R(t)^6 + R(t)^8 \approx 2 R(t)^4$ on the interval $R(t) \in [0, 1]$. Performing this substitution into \eqref{supp_oadyn_ss} results in an approximate solution with the form
\[
R=0, \quad
R \approx \dfrac{\sqrt{2}}{4}  \sqrt{\sqrt{u^2 +16 (u - 1) -16 \left(\dfrac{2 \pi  \sigma_\omega}{V_0}\right)}   - u   }
\]
where $u \equiv \pi K/V_0$. These solutions have respective eigenvalues
\begin{align*}
\lambda_0 &= \dfrac12\left(K -2 \sigma_\omega - \dfrac{V_0}{\pi N}\right),\\
\lambda_1 &= 4 \sigma_\omega - 2 K - \dfrac{\pi K^2 N}{8 V_0} + \dfrac{2 V_0}{\pi N} \\
&+ \dfrac{ N K^2}{8 V_0} \sqrt{\pi^2 + \dfrac{16 \pi V_0}{K N} - \dfrac{16 V_0}{K^2 N}\left(2 \sigma_\omega \pi + \dfrac{V_0}{N}\right)}
\end{align*}
Both the full and approximate solutions undergo a transcritical bifurcation at $V_0 = \pi N (K - 2\sigma_\omega)$. Figure S2 shows the the full and approximate solutions for several values of $K$,  $V_0$, and $\sigma_\omega$, illustrating the transcritical bifurcation between partial synchrony and maximal desynchony.

\begin{figure}
{
\centering
\includegraphics[width=\linewidth]{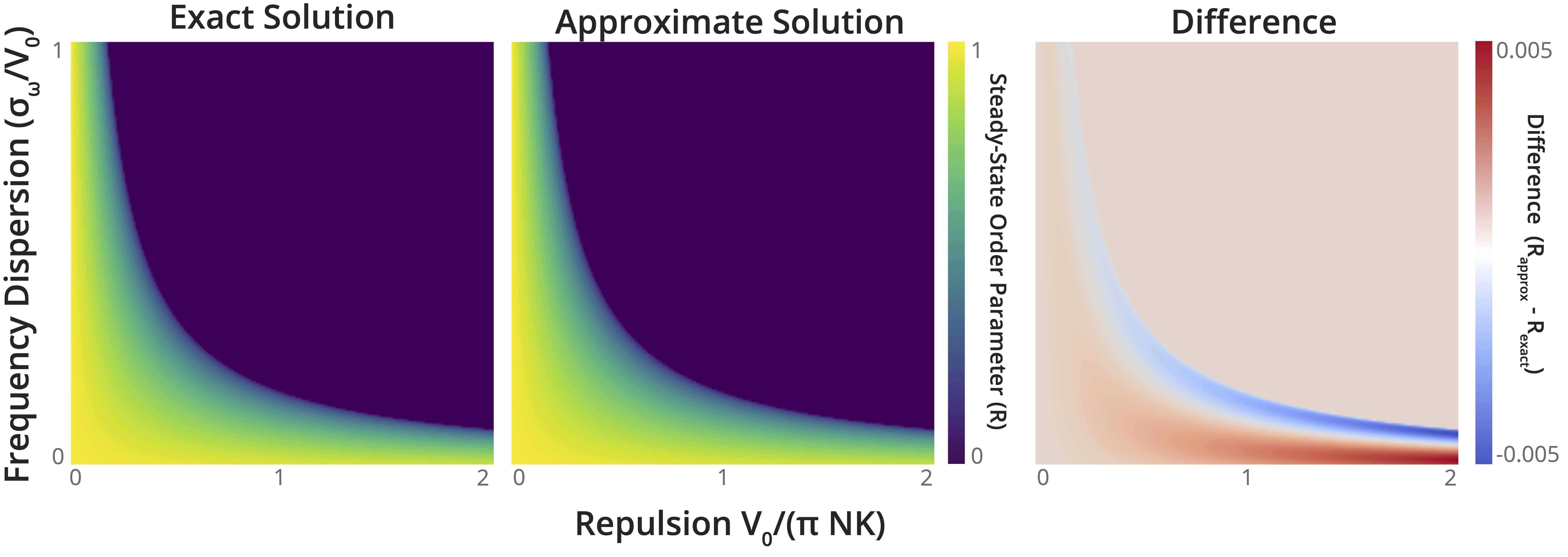}
\caption{
{\bf Exact vs. Approximate solution to the dynamics without noise.} The steady-state solution of the full dynamics, parametrized by $V_0$ and $\sigma_\omega$, for the exact dynamics and those given by the quartic approximate solution. For these plots, $\xi_0 = 0$.
}
\label{approx}
}
\end{figure}

\section{Ott-Antonsen reduction with noise}

\subsection{Rearrangements as noise proportional to the gradient}

We use a heuristic model of rearrangements as a form of noise acting on the reduced-order dynamics of the oscillators in the continuum limit found above. Because the noise force must affect the dynamics of the order parameter $R(t)$, there are restrictions on how it can enter into the dynamical equations for the evolution of the oscillator phase distribution $f(\theta)$. We emphasize that our noise force $\xi$ does not describe Langevin dynamics of individual oscillators. Rather, we assume deterministic microscale dynamics and then derive a continuum model using the Ott-Antonsen ansatz, and we then introduce the stochastic forcing term $\xi$ acting directly on the mean-field dynamics of $R$, in order to "seed" random toppling events.

We start with the hydrodynamic equation for the time evolution of the oscillator distribution $f(\theta)$, as well as the Ott-Antonsen ansatz for the form of $f(\theta)$,
\begin{equation}
\partial_t f + \partial_\theta (v f) = 0, \quad f = \dfrac{1}{2 \pi} \left(1 + \sum_{n=1}^\infty a^n e^{i n \theta} + \text{c.c.}\right).
\label{supp_hydro2}
\end{equation}
We define $v = v_0(\theta, t) + v_\xi(\theta, t) $, where $v_0(\theta, t)$ refers to the forcing on oscillators in the absence of noise, 
\[
v_0 = \omega + \dfrac{K}{2 i} \left( Z e^{-i \theta} - \bar{Z} e^{i \theta} \right)  -  \dfrac{V_0}{N} \partial_\theta f,
\]
and $v_\xi(\theta, t)$ refers to the noise term in the oscillator dynamical equations. In the following subsections, we insert this term and various forms of $v_\xi(\theta, t)$ into the derivation of Section \ref{appendix_oa_no_noise}, in order to determine how various forms of noise affect the dynamics of the order parameter $R(t)$.

{\it Additive case.} One option for the presence of noise in \eqref{supp_hydro} is $v_\xi(\theta, t) = \xi(t)$, such that $v = v_0 + \xi(t)$. In this case, $\dot R(t)$ has identical form as \eqref{supp_oadyn}. The noise solely affects the dynamics of the phase,
\[
\dot \Phi(t) = \omega + \xi(t)
\]

{\it Multiplicative case.} Another option for the presence of noise in \eqref{supp_hydro} is $v_\xi(\theta, t) = f(\theta)^m \xi(t)$, such that $v = v_0 + f(\theta)^m \xi(t)$, where $m > 0$. For the case of $m=1$, \eqref{oa_a} has the form
\begin{align*}
\dot a(t) = -\dfrac{a(t)}{2 \pi}\bigg( 2 \pi i \omega  &- \pi K  + \pi K \abs{a(t)}^2 \\
&+ (V_0/N + 2 i \xi(t)) P(\abs{a(t)})		\bigg)
\end{align*}
with high degree polynomial $P(q) \equiv  1 + q^2 + q^4 + q^6 + q^8$. This corresponds to $\dot R(t)$ having identical form to \eqref{supp_oadyn}, while the phase $\Phi(t)$ has dynamics given by
\[
\dot \Phi(t) = \omega + \dfrac{\xi(t)}{\pi}\left(	1 + R(t)^2 + R(t)^4 + R(t)^6 + R(t)^8	\right)
\]
A similar derivation may be used to show that the dynamics of $R(t)$ are unaffected by multiplicative noise to arbitrary order $m$.

{\it Gradient case.} A final option for the presence of noise in \eqref{supp_hydro} is $v_\xi(\theta, t) = \frac{\partial^m f(\theta)}{\partial \theta^m} \xi(t)$, such that $v = v_0 + \frac{\partial^m f(\theta)}{\partial \theta^m} \xi(t)$, where $m > 0$. For the case of $m=1$, \eqref{oa_a} has the form
\begin{align*}
\dot a(t) = -\dfrac{a(t)}{2 \pi}\bigg(2 \pi i \omega &- \pi K  + \pi K \abs{a(t)}^2 \\
&+ (V_0/N + \xi(t))P(\abs{a(t)})			\bigg)
\end{align*}
with high degree polynomial $P(q) \equiv  1 + q^2 + q^4 + q^6 + q^8$. This corresponds to dynamics given by
\begin{align*}
\dot R(t) =  -\sigma_\omega R(t)  -\dfrac{R(t)}{2 \pi}   \bigg(    \pi K& (R(t)^2 - 1) \\
&+ (V_0/N + \xi(t)) P(R(t))		\bigg) 	\\
\dot \Phi(t) = \omega&
\end{align*}
Thus, we find that the simplest manner in which noise can affect the dynamics of $R(t)$ is through a term proportional to the phase density gradient.

\subsection{Derivation of stochastic differential equation}

Based on our analysis of the dynamical equations above, we assume that the noise enters the dynamical equations through a term of the form $\frac{\partial f(\theta)}{\partial \theta} \xi(t)$. As seen above, this is equivalent to making the substitution $V_0/N \rightarrow (V_0/N + \xi(t))$ in the dynamical equations. Isolating terms in $\dot R(t)$ by their order in $\xi(t)$ reveals that the dynamics have the form of a Stratonovich process,
\begin{equation}
dR = F(R) dt + \xi_0 G(R) dW
\label{strato}
\end{equation}
where $\xi_0$ is a noise amplitude term, and
\begin{align*}
F(R) &=  -\sigma_\omega R(t) -\dfrac{R(t)}{2 \pi}   \bigg(    \pi K (R(t)^2 - 1) + (V_0/N) P(R(t))		\bigg) 	\\
G(R) &= -\dfrac{R(t)}{2 \pi}   P(R(t)).
\end{align*}
with high degree polynomial $P(R(t)) \equiv  1 + R(t)^2 + R(t)^4 + R(t)^6 + R(t)^8$. Note that the first "drift" term is identical to the dynamical equation for the zero-noise limit, \eqref{supp_oadyn}. For this study, we assume a standard Langevin noise term, $\langle \xi(t) \xi(t') \rangle = \xi_0 \delta(t-t')$, making $W$ a Weiner process. Using these definitions, we numerically simulate \eqref{strato} using the Euler–Maruyama method.

\section{Cellular automaton traffic model}

Our cellular automaton model is a modified version of the Nagel–Schreckenberg traffic model \cite{nagel1992cellular,paczuski1996self,nagel1995emergent}. The original model evolves on a one-dimensional ring of $L$ sites, each of which can either be occupied ($1$) or unoccupied ($0$). Each occupied site represents a "car," and cars cannot overlap or overtake one another. In the original model, fluctuations in the speed of one car (due to random braking events, for example) provoke the spontaneous formation of large jams of cars positioned bumper-to-bumper within the circular domain. The jam breaks up only when enough time has elapsed for cars at the head of the jam to accelerate away from the jam. 

To capture our hypothesized mechanism for jamming in the repulsive oscillator model, we modify the Nagel–Schreckenberg model in two key ways: (1) Cars travel in motorcades in which all cars are attracted to the mean location of all cars, even if it is behind them. (2) In high-density regions, "toppling" events occur that are the one-dimensional equivalent of topples in classical sandpile models of criticality \cite{bak1987self}. For each car in the motorcade, if the number of cars directly afore it exceed those directly behind it (or vice versa), then the car has some probability of moving to the nearest unoccupied lattice site. This probability increases as the difference in afore and aft cars increases. Overall, this probabilistic toppling rule causes cars located regions of a jam with a high pressure differential to locally rearrange their positions within the jam more frequently, by analogy to gradient-driven rearrangements among oscillators in our repulsive oscillator simulations.

All together, our cellular automaton model comprises the following steps:
\paragraph{Initialization.} For a given density parameter $\rho$, $M = \text{floor}{(\rho \, L)}$ sites are randomly chosen to have cars at the start of the simulation. A set of random, integer velocities, $\{ v_i \}$, $v_i \in \{0, 1, ..., v_{max}\}$ is assigned to the set of cars. In subsequent timesteps, the locations of all of the cars are then updated according to the following rules:

\paragraph{Iteration.} Within each timestep of the model, the following steps are performed:
\begin{enumerate}
\item {\bf Calculation of the mean location.} The mean location of all $M$ cars, $\bar x$, is calculated. For each car $i$, the number of sites $d_i$ between its location $x_i$ and the mean location $\bar x$ is then calculated. Periodic boundary conditions are assumed; if the forward distance along the circle is more than $L/2$ sites, then the negative backward distance is used. Therefore, $d_i \in [-L/2, L/2]$.
\item {\bf Mean field velocity update.} The velocity of each car, $v_i$, is updated using the following relation: $v_i \leftarrow (1-s)\, v_i + s \, d_i$. The synchronization parameter $s$ is treated as a model parameter; if $s=0$, then the mean location does not affect the cars' velocities at all, but if $s=1$ then the cars always instantaneously update their velocity to their distance from the mean location.
\item {\bf Toppling of large jams.} From the list of $x$ positions, jams are identified as sequences of consecutive $1$ values in the ring of $L$ car positions. These jams are "toppled" according to the following rules:
\begin{enumerate}
\item In each jam, the pressure gradient on each car in the jam, $\partial p_i$, is calculated as the different between the number of cars in behind it in the jam, minus the number of cars in front of it in the jam. For example, the jam corresponding to the sequence $0,1,1,1,1,1,0$ corresponds to a sequence of pressure gradient values $0, -4, -3, 0, 3, 4, 0$. In order to remain physical, a cutoff parameter is imposed, which corresponds to a maximum radius over which to impose the gradient. For example, with a cutoff radius of $2$, the pressure gradient of the jam $0,1,1,1,1,1,0$  is $0, -2, -1, 0, 1, 2, 0$.
\item A random subset is chosen of all the cars that currently exceed a threshold $\abs{\partial p_i} > p_{thresh}$. Cars with larger pressure gradients have a larger probability of being chosen. We note that the value of $p_{thresh}$ sets the average frequency and amplitude of toppling events, but does not otherwise affect the dynamics.
\item Among all cars in the random subset, the velocity of each car in the jam is increased by an amount proportional to the pressure gradient, $v_i \leftarrow v_i + f \partial p_i$, where $f$ is the amplitude of the rearrangement force.
\end{enumerate}
\item {\bf Acceleration and maximum velocity.} The speeds of all cars are increased by one, $v_i \leftarrow v_i + 1$. Any cars that have a velocity greater than the maximum velocity are reset to the maximum velocity, $v_i \leftarrow \min\{v_i, v_{max}\}$
\item {\bf Braking to avoid collisions.} The distance between each car, and the next car in front of it, is calculated in order to produce a set of spacings $\eta_i$. For cars that have a velocity larger than the spacing, the velocity is reset to equal the spacing. $v_i \leftarrow \min\{\eta_i, v_{max} \}$. Following the original Nagel–Schreckenberg model, among the cars that brake, a random subset is chosen to "overbrake" by $1$ unit, $v^{samp}_i \leftarrow v^{samp}_i - 1$, with probability $p_{brake}$.
\item {\bf Position update.} The positions of all cars are updated using the final value of the velocity, $x_i \leftarrow x_i + v_i$
\end{enumerate}

The primary parameters that govern the behavior of the model are the density $\rho$, the synchronization parameter $s$, the magnitude of the critical force $p_{thresh}$, and the fraction of cars that overbrake, $p_{brake}$. Consistent with the original Nagel–Schreckenberg, if $p_{thresh}$ approaches $1$, the size and durations of jams increases.

\section{Numerical stability of Avalanches}
\label{numerics}

\begin{figure}
{
\centering
\includegraphics[width=\linewidth]{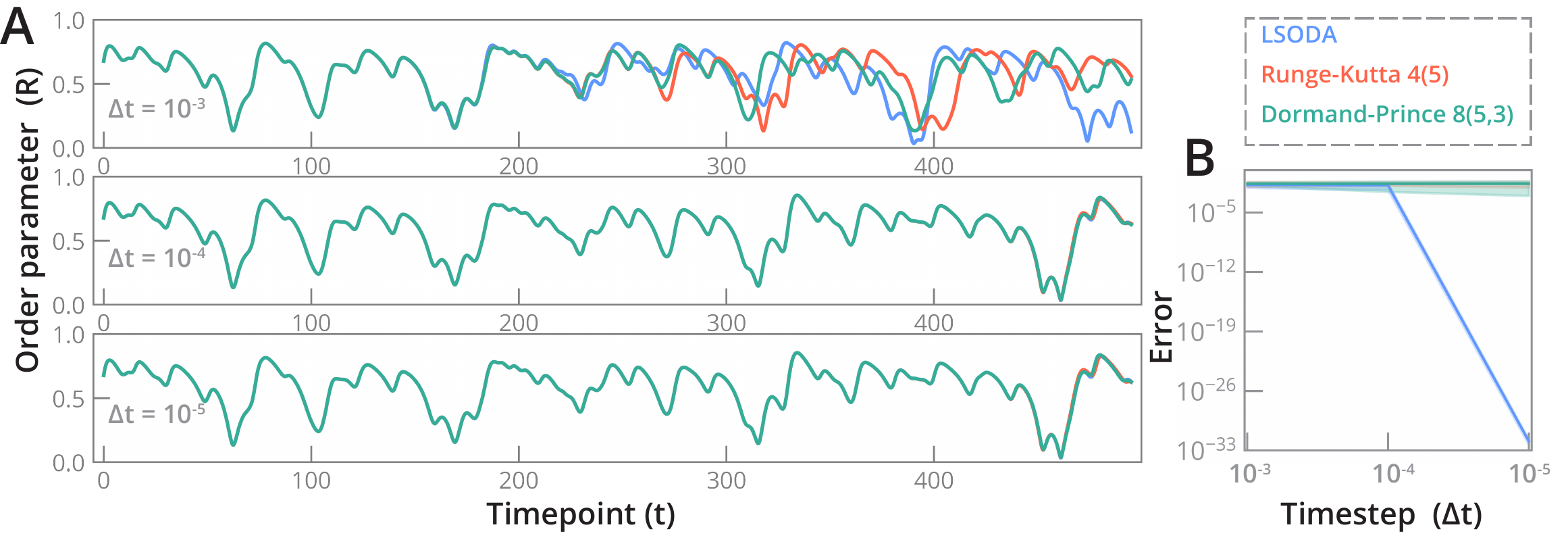}
\caption{
{\bf Numerical robustness of avalanche results.} (A) Example trajectories of the Kuramoto order parameter, for different integration methods (colors) and for different maximum timesteps (panels). Trajectories for the LSODA and Runge-Kutta methods are obscured by the Dormand-Prince trajectory in the lower two panels, indicating strong agreement. (B) Mean square error between the observed trajectories, and a fiducial trajectory calculated with maximum timestep of $\Delta t = 10^{-7}$. Error bars correspond to replicates with different initial conditions. For this panel, $N=40$, $K = 0.4$, $V_0 = 0.8$, $\sigma_\omega=0.2$.
}
\label{numeric}
}
\end{figure}

Due to the separation of timescales between synchronization and repulsion in our system, we consider the degree to which our observed dynamics are robust to details of the numerical integration scheme, in order to rule out numerical artifacts. We compute an expensive fiducial trajectory using an integration timestep of $\Delta t = 10^{-7}$, and we use it to assess the accuracy of trajectories generated with different numerical integrators and timesteps (Figure \ref{numeric}). Across three different integrators (two fixed step, one variable-step), we find that long-timescale simulations are consistent as long as the maximum timestep is less than $10^{-4}$, and that the accuracy is highest for LSODA.

\clearpage
\bibliography{avalanche_cites} 
\bibliographystyle{naturemag}

\end{document}